\documentstyle[12pt]{article}
\normalsize

\def\br{\begin{thebibliography}{abc}}
\def\er{\end{thebibliography}}

\newcommand{\bc}{\begin{center}}
\newcommand{\ec}{\end{center}}
\begin{document}
\newcommand{\be}{\begin{equation}}
\newcommand{\ee}{\end{equation}}
\newcommand{\bea}{\begin{eqnarray}}
\newcommand{\eea}{\end{eqnarray}}


\baselineskip=24pt




\begin{center}\bf METAL ADATOM INDUCED CORRUGATION OF Cu(001)\end{center}

Wei Li and Gianfranco Vidali

Department of Physics, Syracuse University, Syracuse, NY 13244  USA
\bigskip

\noindent ABSTRACT

\medskip\noindent
We report the discovery of a metal adatom induced corrugation of Cu(001)
as probed by atom beam scattering (ABS). At Pb or Bi coverages of 0.05
(fraction of Cu layer), and while a (1$\times$1) LEED pattern is still
observed, the ABS diffraction pattern from \underbar{uncovered} Cu(001) areas
has changed considerably from that of clean Cu(001). We show that these
uncovered areas have become up to ten times more corrugated than the clean
Cu(001).  We performed several carefully designed experiments to test whether
our data could be interpreted as a result of scattering from adsorbate islands
or from a lattice of randomly occupied sites. Instead, we ascribe the nature of
this metal adatom induced corrugation to the redistribution of the Cu(001)
surface charge density due to the adsorption of Pb or Bi atoms. No effect is
seen following adsorption of weakly chemisorbed adatoms, such as Hg. A model
consistent with all our observations is presented.

\break
\newpage
\noindent
I. INTRODUCTION

The study of structural changes of a surface upon adsorption of other atoms is
not just a very challenging problem {\it per se}, but it is important in
understanding how heteroepitaxy proceeds. Structural changes of metal and
semiconductor surfaces upon adsorption of non-metal elements have been studied
extensively \cite{1}. As to structural changes of metal surfaces upon
adsorption of metal atoms, most of the attention has been devoted to
substantial substrate reconstructions due to adsorption of alkali metal
atoms \cite{2}. Other restructuring processes studied recently include surface
layer
relaxation and faceting \cite{3}.

\noindent
Here we report a qualitatively different type of surface restructuring induced
by metal atoms adsorption. We find that adsorption at low coverage
of Pb or Bi atoms on Cu(001) produces a change of the outer edges of the
substrate electron density distribution. The key point is that these changes
are not as dramatic as the ones observed for alkali metal adsorption on most
metal surfaces \cite{2}, yet they
are significant and they might be used as a test ground for theoretical models
of adsorbate induced restructuring \cite{4}.

We employed the technique of Atom Beam Scattering (ABS) to measure such
changes. We briefly recall that ABS directly probes the electron charge density
profile of a surface \cite{5}. At typical helium beam energies employed,
surface
electron densities of the order of 10$^{-4}$e$\AA^{-3}$  are probed; these
densities are usually found at about 3 $\AA$ from the plane of the nuclei.
Compact metal surfaces, such as Cu(001), have extremely flat profiles (in
directions parallel to the substrate) and typically no helium beam diffraction
peaks are detected \cite{6} except for one case which was attributed to surface
impurities \cite{six}. Such flatness can be translated in surface electron
charge density corrugation amplitudes of less than 0.01 $\AA$.

A communication about these experiments with Pb depositions has been given
recently \cite{7}; here we report on additional experiments with Bi and Hg and
on
details of our simulations.

\noindent
II.EXPERIMENTAL

Our experiment was carried out in a UHV system equipped with He beam
scattering, LEED and  Auger electron spectroscopy \cite{8,eight}. A helium beam
with of
18.4 meV energy and about 1\% (FWHM) velocity distribution was used in this
study.  The thoroughly desulfurized Cu(001) sample was prepared by repeated
cycles of Ar ion sputtering and annealing at 580$^{\circ}$C prior to each run.
The well prepared Cu(001) sample yielded more than 50\% of the incident beam in
the specular reflection and the specular FWHM was close to the detector
resolution, currently set at 0.7$^{\circ}$.  The sample temperature was changed
between 120 K and 873 K. Pb or Bi of 99.999\% purity was deposited by using a
liquid nitrogen shielded Knudsen evaporation source with the Cu surface at
either 150 K or 410 K. Coverages were calibrated with both the ABS real-time
deposition curve \cite{8,eight} and Auger signals.

\noindent
II. RESULTS

The result of this study is the observation of the emergence of a diffraction
pattern with the two dimensional periodicity of Cu(001) when a small amount of
Pb or Bi is deposited.   In a typical experiment, we proceed as follows.
First, we check that no diffraction peaks are present from a clean Cu(001)
surface but the specular peak, in agreement  with previous
investigations \cite{6}. Then Bi is deposited on Cu(001) at 410 K.  At a
coverage
of 0.05, fraction of the Cu layer, we observe the emergence of diffraction
peaks along azimuth $\Phi$=0$^{\circ}$ ($<$110$>$ direction of Cu) and
$\Phi$=45$^{\circ}$, which are coincident with the (0,-1) and (-1,-1) peaks of
Cu(001), respectively. Typical detector scans as a function of Bi coverage are
shown in Fig.1. Similar results were
obtained using Pb \cite{7}. In Fig. 2 we show the intensity of the (-1,-1)
Cu(001)
diffraction peak as a function of Pb coverage. Systematic polar and azimuthal
angle scans were done; no non-integer order ABS diffraction peaks or other
features (as due to scattering from isolated adatoms) were seen within the
sensitivity of the apparatus (10$^{-4}$ of specular peak). A small background
enhancement due to diffuse scattering was detected. Throughout these
experiments the LEED pattern always showed a (1$\times$1) phase with slight
intensity increase of integer spots and an increased background.

To show the diffraction peaks are indeed from the still uncovered parts of the
Cu(001) surface, we performed the following experiment. A small amount
of Pb was put on the surface and diffraction peaks with the symmetry of Cu(001)
were measured, Fig 3a. Then a layer of Hg was deposited with the substrate at
250 K - in this condition no more than one layer is absorbed \cite{9}. The
diffraction peaks have disappeared, Fig 3b, because Hg covered the previously
unexposed Cu(001) areas. The Hg layer is desorbed and the peaks reappear, see
Fig.3c.

\noindent
III. DISCUSSION

The clear observation of this study is the emergence of diffraction peaks with
the Cu(001) periodicity. The missing of
non-integer order diffraction  peaks (in both ABS and LEED) indicates
that adsorbed Pb atoms do not form a dilute ordered phase. Furthermore, the
size of a Pb atom (3.6$\AA$) prevents Pb atoms
from forming any ordered phase with a lattice constant of Cu(001) (2.55$\AA$).
 From ABS measurements there is no indication of island formation and it is
well
known that Bi adatoms have repulsive interactions \cite{10}.

To explain these observations we present the following model. Above a threshold
coverage of 0.05  adatoms
induce a corrugation of the Cu(001) surface. Diffraction from exposed but
affected areas of copper start to develop (Fig.2). As the coverage increases,
more
Cu(001) areas
adjacent to Pb atoms are affected and the signal is proportional to $\Theta^2$;
eventually the coverage becomes large enough
that Cu(001) unexposed areas (proportional to $\Theta$) start to shrink.
Therefore we have a relation of peak intensity I$_G$ vs. coverage $\Theta$:
\be
I_G=I_0(\Theta-\Theta_0)^2(1-C(\Theta-\Theta_0))^2
\ee
\noindent From fitting we obtain C=3.8. The diffraction peaks disappear at
coverage $\Theta_s=\frac{1}{C}$, which corresponds to an average distance
between Pb atoms of 5.6$\AA$. $\Theta_0$ is the initial coverage offset.
The fitting curve is displayed in Fig.2.


Diffraction peaks with the periodicity of Cu(001) appear because of a transfer
of electron charge density. In other words, Pb or Bi adsorption causes the
electron
charge density to accumulate on the Cu atoms giving rise to a more pronounced
profile.  No effect is seen for Hg adsorption, presumedly
because Hg is very weakly chemisorbed \cite{9} and thus does not affect the Cu
substrate \cite{7}.

If Pb or Bi atoms adsorb on commensurate sites but randomly, calculations show
that diffraction peaks with the commensuration of the substrate lattice appear
\cite{13}. A very weak ( 10$^{-5}$ of specular intensity) diffraction peak due
to a lattice gas of CO on Ni(100) was detected up to coverages of 0.3
\cite{14}. The following analysis shows that the lattice gas effect gives only
a minor contributor to the diffraction peaks we measure. Using a hard wall
model in the Eikonal approximation, we calculated the scattering of a He atom
from a chain of Pb atoms adsorbed at random commensurate sites of a 300 atom
long Cu chain (see Fig. 4a). At the low coverage considered here
($\Theta$=0.14), each Pb atom was modeled as a Gaussian bump of
4$\AA$ height and 6.5$\AA$ wide (FWHM). These parameters were obtained from a
calculation of the charge density of a Pb adatom \cite{15}. A similar method
was used in
Ref.\cite{16} to model the scattering of He from a collection of Na atoms. For
the kinematic conditions used in the experiment, the first diffraction peak
from a 1D lattice model is 3$\times$10$^{-3}$ of the specular peak for an
average of 20 random configurations. Now we introduce a
{\underbar small} corrugation for the substrate by using a function of $\xi
cos\left({2\pi \over{a}} x\right)$ (see Fig 4b), here a=2.55$\AA$ is Cu(001)
lattice constant. Using $\xi$=0.04$\AA$ obtained from a 2D hard wall fitting to
the experimental data \cite{7}, the same diffraction
peak intensity jumps to 5$\times$10$^{-2}$ of the specular peak, which is close
to experimental
data.  Although this corrugation is much less than the ones found for open
reconstructed surfaces, it is at least ten times bigger than the one for a
clean Cu(001).   Clearly the lattice gas effect is small.

\noindent
IV. CONCLUSIONS

We have measured diffraction peaks with the symmetry of Cu(001) when small
amounts of Pb or Bi were deposited. We attribute
the emergence of these peaks to  a rearrangement of the outer edges of the
Cu(001) electron charge density. We have also shown that other effects, such as
scattering from a lattice gas, cannot account for the magnitude of the
observed phenomena. As to the detailed mechanism that causes the
reconstruction, other types of probes and theoretical calculations are
necessary to obtain more information. At this point we speculate that Pb or Bi
adsorption brings about an
instability of the relaxed clean Cu(001) surface. The rearrangement of
electrons, although tiny in absolute values, might be connected to a change of
an sp-like surface band just below the Fermi level \cite{17}. Indeed, ABS is
most
sensitive to s or p electrons, since those are the ones spilling most into the
vacuum.

\noindent
ACKNOWLEDGMENTS

This work was supported by NSF grant DMR 8802512.


\newpage
\noindent
FIGURE CAPTIONS

Figure 1. (0,-1) peak profiles at different Bi coverages;
$\theta_i$=60$^{\circ}$. Substrate temperature is 410 K.   Solid lines are fits
to a Gaussian function with a linear background subtraction.

Figure 2. (-1,-1) diffraction peak intensity vs. Pb coverage. Conditions as in
Fig.1. The solid line is a fit to Eq.(1).

Figure 3. (-1,-1) peak profile: a) after depositing Pb ($\Theta$=0.13); b)
after depositing a layer of Hg on top of surface as in a); c) after desorbing
Hg. Substrate temperature is 250 K.  $\theta_i$=60$^{\circ}$.  Solid lines are
fits to a Gaussian function with a linear background subtraction.

Figure 4. Sketch of 1D model. a) Cu chain is flat and Pb atoms are adsorbed
randomly at commensurate sites; b) as in a) but now the Cu chain has a small
corrugation.

\end{document}